# Density matrix approach for quantum free-electron lasers

H. Fares[1,2,*], G. R. M. Robb[3], N. Piovella[4]

[1]INFN-LNF - Via Enrico Fermi, 40, I-00044, Frascati (Roma), Italy
[2]Department of Physics, Faculty of Science, Assiut University, Assiut 71516, Egypt
[3]Department of Physics & SUPA, University of Strathclyde, Glasgow, G4 0NG Scotland, United Kingdom
[4]Dipartimento di Fisica "Aldo Pontremoli", Università degli Studi di Milano & INFN, Sezione di Milano, via Celoria 16, I-20133 Milano, Italy
* hesham.fares@lnf.infn.it
Tel:+3472871745. Office:+694032254.

**Abstract**

The density matrix in the Lindblad form is used to describe the behavior of the Free-Electron Laser (FEL) operating in a quantum regime. The detrimental effects of the spontaneous emission on coherent FEL operation are taken into account. It is shown that the density matrix formalism provides a simple method to describe the dynamics of electrons and radiation field in the quantum FEL process. In this work, further insights on the key dynamic parameters (e.g., electron populations, bunching factor, radiation power) are presented. We also derive a simple differential equation that describes the evolution of the radiated power in the linear regime. It is confirmed that the essential results of this work agree with those predicted by a discrete Wigner approach at practical conditions for efficient operation of quantum FELs.

**1. Introduction**

The Free−Electron Laser (FEL) operating in a quantum regime, the so-called quantum FEL (QFEL), has been proposed as a potential compact, tunable, near monochromatic, hard x−ray source [1,2]. The characteristics of a QFEL are crucial for many demanding applications, such as medical, commercial, and academic research applications. The quantum regime of FEL is realized when the induced momentum spread of the electron $\delta p_z = mc\gamma\rho_{FEL}$ is smaller than the photon momentum $\hbar k$ where $\rho_{FEL}$ is the FEL parameter and $\gamma$ is the electron energy [3,4]. Then, in order to identify the regime of FEL operation, i.e., whether it is classical or quantum, a dimensionless parameter $\bar{\rho} = \delta p_z/\hbar k$ has been introduced [3,4]. It has been noted that the quantum regime (i.e., when $\bar{\rho} < 1$) is more easily realizable at higher photon energies [1,2]. In the quantum regime, an electron can be represented as a two−level system, and each electron emits at most one photon when the saturation takes place. The fundamental characteristic of the QFEL is an extremely narrow spectrum due to the discreteness of momentum exchange. On the other hand, in the classical regime of FEL (i.e., when $\bar{\rho} \gg 1$), numerous transitions between several momentum states occur and then a multi−frequency spectrum is observed [5].

In the FEL operating in the quantum regime, the spontaneous emission represents a loss mechanism and can significantly hinder the coherent FEL emission. The frequency of the spontaneously radiated photon depends on the angle of the emitted photon with respect to the electron beam direction. Therefore, the spontaneous emission is characterized by a broadband spectral range. In Refs. [6,7], a quantum−mechanical model based on a continuous Wigner function has been developed for describing the QFEL operation including spontaneous emission. In Ref. [7], the authors determine the condition at which the effect of spontaneous emission is negligible. It has also been demonstrated that the inclusion of the broadband frequency of the spontaneous emission is insignificant. Then, the spontaneous emission can be assumed monochromatic whereas it is almost emitted in the forward direction of the electron beam. In Ref. [8], a model based on discrete Wigner function has been developed for describing the coherent radiation of QFEL, and recently extended in Ref. [9] to include the spontaneous emission effect. In this model, the electron momentum is assumed to be a discrete variable consistent with the quantum nature of the emitted radiation. Then, the approach of the discrete Wigner function used in Ref. [9] is more exact than that based on the continuous Wigner function used in Ref. [7]. It is noticed that in the discrete Wigner model [9], the spontaneous emission is also described as monochromatic photons emitted randomly by electrons as assumed in Ref. [7].

In this paper, a simplified model based on the density matrix formalism is presented for describing the QFEL interaction. In this model, the dynamics of electrons undergoing spontaneous emission are described using the Lindblad master equation for the density matrix [10,11]. We confirm the validity of the density matrix model in the practical regime of QFEL operation at low or even moderate spontaneous emission rates. Approximately, the moderate spontaneous emission rate refers to that which reduces the coherent intensity to about half of its maximum value. Although the regime of high spontaneous emission rate is impractical, we report on the invalidation of the density matrix approach in this regime. A quantitative criterion for applying the density matrix treatment is presented. In this work, we show that the density matrix model presents effective tools for further understanding of the dynamics of electron and radiation fields in the QFEL.

This paper is organized as follows. In section 2, the discrete Wigner model of QFEL shown in [9], as a benchmark model, is reviewed. In section 3, we present a model based on the density matrix formalism for the QFEL. Similar to the Wigner model, a system of coupled equations for describing the evolution of the QFEL process including the spontaneous emission is derived. Expressions for the density matrix elements that describe the dynamics of electrons are obtained using a master equation in the Lindblad form. In section 4, numerical examples are given providing a further understanding of the dynamics of electrons in the QFEL under the influence of the spontaneous emission. By comparing the results of the density matrix and the discrete Wigner approaches, we address the question of the validity of the density matrix model introduced in this work. Section 5 is devoted to conclusions.

## 2. Discrete Wigner model

In this section, we review the basic results of the discrete Wigner function approach proposed for QFEL involving the spontaneous emission, as described in Ref. [9]. Then, a comparison with the results of a density matrix—based model, newly introduced in this work, can be made.

In the quantum theory of FELs [8], the ponderomotive electron phase $\theta=(k+k_w)z-kct$ is assumed to be a periodic variable in $(0,2\pi]$ where $k$ and $k_w$ are the wave number of radiation and wiggler, respectively. This hypothesis assures that the conjugate momentum variable $p$ is discrete. A scaled momentum representing the relative electron momentum in unites of $\hbar k$ is $p=mc\,(\gamma-\gamma_0)/\hbar k$ where $\gamma$ and $\gamma_0$ are the instantaneous and initial electron energies in units of $mc^2$, respectively. Accordingly, a $\theta$ —periodic electron state $|\Psi(\bar{z},\theta)\rangle$ is expanded in terms of momentum eigenstates $|n\rangle$ as [8]

$$|\Psi(\bar{z},\theta)\rangle=\frac{1}{\sqrt{2\pi}}\sum_{n=-\infty}^{n=\infty}c_n(\bar{z})\,|n\rangle,\quad \langle\theta|n\rangle=e^{in\theta}. \tag{1}$$

In Eq. (1), $\bar{z}=z/L_g$ is a normalized distance where $L_g=\lambda_w/4\pi\rho_{FEL}$ is the gain length and $\lambda_w$ is the wiggler period. $|c_n|^2$ is the probability of finding an electron in a momentum state $n$. The eigenstates $|n\rangle$ satisfy the eigenvalue equation

$$\hat{p}|n\rangle=n|n\rangle,\ \ \text{where}\,\hat{p}=-i\frac{\partial}{\partial\theta}. \tag{2}$$

The operators $\hat{\theta}\equiv\theta$ and $\hat{p}$ satisfy the commutation relation $[\hat{\theta},\hat{p}]=i$.

The electron dynamics is described by a Schrodinger—like equation [8]

$$i\frac{\partial|\Psi(\bar{z},\theta)\rangle}{\partial\bar{z}}=\hat{H}|\Psi(\bar{z},\theta)\rangle, \tag{3}$$

where $\hat{H}$ is the single—electron Hamiltonian and is given as

$$\hat{H}=\hat{H}_0+\hat{H}_{\text{int}}=\frac{\hat{p}^2}{2\bar{\rho}}-i\bar{\rho}\big(Ae^{i\theta}-c.c.\big). \tag{4}$$

In Eq. (4), $\hat{H}_0=\hat{p}^2/2\bar{\rho}$ is the unperturbed Hamiltonian and $\hat{H}_{\text{int}}=-i\bar{\rho}\big(Ae^{i\theta}-c.c.\big)$ is the interaction Hamiltonian.

On the basis of the above discussion, an approach based on a Wigner distribution function with periodic boundaries in $\theta$ has been developed to formulate the quantum theory for FELs [9]. In this approach, the system of coupled equations that describes the FEL including spontaneous emission is written as [9]

$$\begin{aligned}\frac{\partial w_s(\bar{z},\theta)}{\partial\bar{z}}&+\frac{s}{\bar{\rho}}\frac{\partial w_s(\bar{z},\theta)}{\partial\theta}\\&=\bar{\rho}\big(Ae^{i\theta}+c.c.\big)\big\{w_{s+1/2}(\bar{z},\theta)-w_{s-1/2}(\bar{z},\theta)\big\}+\frac{\beta}{\bar{\rho}}\{w_{s+1}(\bar{z},\theta)-w_s(\bar{z},\theta)\},\end{aligned} \tag{5}$$

$$\frac{dA}{d\bar{z}}=\sum_{m=-\infty}^{\infty}\int_{-\pi}^{\pi}w_{m+1/2}(\bar{z},\theta)e^{-i\theta}d\theta+i\delta A. \tag{6}$$

In Eq. (5), $w_s(\bar{z},\theta)$ represents two types of Wigner functions where $s=m$ or $s=m+1/2$. $\beta=\alpha a_w^2 mc\gamma_r/6\hbar k$ is the scaled spontaneous emission rate where $\alpha$ is the fine structure constant and $a_w=eB_w/k_w mc$ is the wiggler parameter. $A$ is a scaled complex amplitude of the radiation field whereas the photon number emitted by each electron is $\bar{\rho}|A|^2$ [3]. In Eq. (6), $\delta=(\gamma_r-\gamma_0)/\rho_{FEL}\gamma_0$ is the detuning parameter and $\gamma_r$ is the resonant energy.

Since $w_s(\bar{z},\theta)$ is periodic in $\theta$, it can be represented as a Fourier series in the form of [8,9]

$$w_s(\bar{z},\theta)=\frac{1}{2\pi}\sum_{n=-\infty}^{\infty} w_s^n(\bar{z})e^{in\theta}. \tag{7}$$

The Fourier components $w_s^n(\bar{z})$ are associated to $c_m(\bar{z})$ of the wave function $\Psi$ where $w_m^{2n}=c_{m+n}^*c_{m-n}$ and $w_{m+1/2}^{2n+1}=c_{m+n+1}^*c_{m-n}$ [8]. Then, $w_m^0=|c_m|^2$ is the population of the $m$ −th momentum state and $w_{m+1/2}^1=c_{m+1}^*c_m$ is the $m$ −th bunching component.

In the quantum regime, $\bar{\rho}\ll 1,$ a two level system is considered where the electron transition occurs between two adjacent momentum states, $m=0$ and $m=-1$.

Substituting Eq. (7) in Eq. (5), considering the terms with $s=0$, $s=-1$, and $s=-1/2$, neglecting the higher components $w_0^2$ and $w_{-1}^2$, and defining the populations parameters $P_0=w_0^0$ and $P_{-1}=w_{-1}^0$ and the bunching parameter $B=w_{-1/2}^1$ (i.e., $B^*=w_{-1/2}^{-1}$), we finally can write the coupled equations for the QFEL as [9]

$$\frac{dP_0}{d\acute{z}}=-\left(\acute{A}\acute{B}^*+c.c.\right)-DP_0, \tag{8}$$

$$\frac{dP_{-1}}{d\acute{z}}=\left(\acute{A}\acute{B}^*+c.c.\right)+D(P_0-P_{-1}), \tag{9}$$

$$\frac{d\acute{B}}{d\acute{z}}=-\left(i\acute{\delta}+D\right)\acute{B}+\acute{A}(P_0-P_{-1}), \tag{10}$$

$$\frac{d\acute{A}}{d\acute{z}}=\acute{B}. \tag{11}$$

In Eqs. (8)−(11), we define the new variables $\acute{z}=\sqrt{\bar{\rho}}\bar{z}$, $\acute{A}=\sqrt{\bar{\rho}}Ae^{-i\delta\bar{z}}$, $\acute{B}=Be^{-i\delta\bar{z}}$, $\acute{\delta}=[\delta-(1/2\bar{\rho})]/\sqrt{\bar{\rho}}$, and $D=\beta/\bar{\rho}^{3/2}$.

In Ref. [9], by solving Eqs. (8)—(11) numerically, many simulations have been carried out to investigate the detrimental effect of spontaneous emission on the FEL operating in the quantum regime. It has been shown that the effect of spontaneous emission is negligible when $D\lesssim 0.07$ in agreement with the results of Ref. [7].

**3. Density matrix model**

In this section, using the density matrix formulation, coupled differential equations for describing the behavior of QFELs are presented. Here, the dynamics of the electrons is described by the master equation in the Lindblad form [10,11]. From the master equation, expressions for the elements of the density matrix operator $\rho$ are obtained. It is instructive to recall that the diagonal elements $\rho_{n,n}$ represent the probability of finding an electron in a particular $n$ −th state, while the off diagonal terms $\rho_{n,m}$ represent the degree of coherence. In this work, for convenience, the density matrix $\rho$ is designated by a subscript $n, m$ (not $nm$). For a two−level system, one can realize the correspondences between the density matrix elements and Wigner function components where $\rho_{0,0} \equiv P_0$ (i.e., $\rho_{-1,-1} \equiv P_{-1}$) and $\rho_{0,-1} \equiv B$. These correspondences allow a direct comparison between the results of the density matrix and the discrete Wigner models.

For a two−level system, the elements of the density matrix $\rho$ are obtained from the Lindblad master equation that takes the form [10]

$$\frac{d\rho(\bar{z})}{d\bar{z}} = -i\left[\hat{H}, \rho\right] - \frac{\beta}{2\bar{\rho}}(\sigma_-\sigma_+\rho + \rho\sigma_-\sigma_+ - 2\sigma_+\rho\sigma_-), \tag{12}$$

where $\sigma_+$ and $\sigma_-$ are the emission and absorption operators, respectively. In the master equation, the spontaneous emission is expressed by the second term on the right−hand side, termed the dissipative term. Notice that the rate of spontaneous emission $\beta/\bar{\rho}$ shown in the dissipative term is implied considering the correspondence between $w_s$ in Eq. (5) and $\rho$ in Eq. (12).

In a 2−dimentional Hilbert space, the momentum states $|0\rangle$ and $|-1\rangle$ can be written as

$$|0\rangle = \begin{pmatrix}1\\0\end{pmatrix}, \quad |-1\rangle = \begin{pmatrix}0\\1\end{pmatrix}. \tag{13}$$

Then, $\sigma_+$ and $\sigma_-$ become

$$\sigma_+ = \begin{pmatrix}0 & 0\\1 & 0\end{pmatrix}, \sigma_- = \begin{pmatrix}0 & 1\\0 & 0\end{pmatrix}, \tag{14}$$

where the relations $\sigma_+|0\rangle = |-1\rangle$ and $\sigma_-|-1\rangle = |0\rangle$ are satisfied. Using Eq. (14), the dissipative term in Eq. (12) is

$$\frac{\beta}{2\bar{\rho}}(\sigma_-\sigma_+\rho + \rho\sigma_-\sigma_+ - 2\sigma_+\rho\sigma_-) = \frac{\beta}{\bar{\rho}}\begin{pmatrix}\rho_{0,0} & \rho_{0,-1}/2\\ \rho_{-1,0}/2 & -\rho_{0,0}\end{pmatrix}. \tag{15}$$

Now, we treat the term $-i\left[\hat{H}, \rho\right]$ in Eq. (12). To simplify our notations, we will first carry out our analysis for unknown momentum states, $n$ and $m$. In a later step, we will replace $n$ and $m$ by 0 and $-1$, respectively.

Since $\hat{H} = \hat{H}_0 + \hat{H}_{\text{int}}$, the term $-i\left[\hat{H}, \rho\right]$ can be rewritten as

$$-i\left[\hat{H}, \rho\right] = -i\left\{\left(\hat{H}_0\rho - \rho\hat{H}_0\right) + \left(\hat{H}_{\text{int}}\rho - \rho\hat{H}_{\text{int}}\right)\right\}. \tag{16}$$

In Eq. (16), the interchange term of the principal Hamiltonian $\hat{H}_0 = \hat{p}^2/2\bar{\rho}$ is

$$-i\langle n|(H_0\rho - \rho H_0)|m\rangle = \frac{i}{2\bar{\rho}}(m^2 - n^2)\,\rho_{n,m}, \tag{17}$$

while the interchange term of the interaction Hamiltonian is

$$\begin{aligned}-i\langle n|\left(\hat{H}_{\text{int}}\rho - \rho\hat{H}_{\text{int}}\right)|m\rangle &= -i\left\{\langle n|\hat{H}_{\text{int}}\rho|m\rangle - \langle n|\rho\hat{H}_{\text{int}}|m\rangle\right\},\\ &= -i\sum_{\ell}\left\{\langle n|\hat{H}_{\text{int}}|\ell\rangle\langle \ell|\rho|m\rangle - \langle n|\rho|\ell\rangle\langle \ell|\hat{H}_{\text{int}}|m\rangle\right\}.\end{aligned} \tag{18}$$

Using the relation of $\hat{H}_{\text{int}} = -i\bar{\rho}\left(Ae^{i\theta} + c.c.\right)$, one can write

$$\langle n|\hat{H}_{\text{int}}|m\rangle = \langle n|-i\bar{\rho}\left(Ae^{i\theta} + c.c.\right)|m\rangle = -i\bar{\rho}(A_{nm} + A^*_{nm}). \tag{19}$$

where $A_{nm} = \langle n|Ae^{i\theta}|m\rangle$ and its complex conjugate is $A^*_{nm} = \langle n|A^* e^{-i\theta}|m\rangle$. $A_{nm}$ is an off−diagonal element that represents the coupling efficiency between the radiation mode and the electron wave. We can assume that the amplitude of the field $A$ is almost constant over one ponderomotive period of an electron wave (i.e., equivalently, over one radiation wavelength when $\theta \in (0,2\pi]$ ). Hence, $A_{nm} = A\langle n \mid e^{i\theta} \mid m\rangle = A\delta_{n,m+1}$ and $A^*_{nm} = A^*\langle n|e^{-i\theta}|m\rangle = A^*\delta_{n,m-1}$.

Using Eqs. (16)−(19), we get

$$-i\langle n|[\hat{H},\rho]|m\rangle = i\frac{(m^2-n^2)}{2\bar{\rho}}\rho_{n,m} + \bar{\rho}\left[A\left(\rho_{n,m+1}-\rho_{n-1,m}\right) + A^*\left(\rho_{n+1,m}-\rho_{n,m-1}\right)\right] \tag{20}$$

Using Eqs. (15) and (20) with Eq. (12) and assuming the initial state is $n = 0$ and the final state is $m = -1$, we get

$$\frac{d\rho_{0,0}}{d\bar{z}} = -\bar{\rho}\left(A^*\rho_{0,-1} + A\rho^*_{0,-1}\right) - \frac{\beta}{\bar{\rho}}\rho_{0,0}, \tag{21}$$

$$\frac{d\rho_{-1,-1}}{d\bar{z}} = \bar{\rho}\left(A^*\rho_{0,-1} + A\rho^*_{0,-1}\right) + \frac{\beta}{\bar{\rho}}\rho_{0,0}, \tag{22}$$

$$\frac{d\rho_{0,-1}(\bar{z})}{d\bar{z}} = \left(\frac{i}{2\bar{\rho}} - \frac{\beta}{2\bar{\rho}}\right)\rho_{0,-1}(\bar{z}) + \bar{\rho}\left[\rho_{0,0}(\bar{z}) - \rho_{-1,-1}(\bar{z})\right]A. \tag{23}$$

Using the density matrix formulations, the expectation value of an arbitrary operator $\mathcal{R}$ is

$$\begin{aligned}\langle\mathcal{R}\rangle = \mathrm{Tr}\{\rho\mathcal{R}\} &= \sum_n \langle n|\rho\mathcal{R}|n\rangle \\ &= \sum_n\sum_m \langle n|\rho|m\rangle\langle m|\mathcal{R}|n\rangle.\end{aligned} \tag{24}$$

The normalized radiation amplitude $A$ is classically determined by the bunching parameter $b = (1/N)\sum_{j=1}^{N} e^{-i\theta_j}$ [12] where $dA/d\bar{z} = b + i\delta A$. Then, according to Eq. (24), $A$ is given quantum mechanically by

$$\frac{dA(\bar{z})}{d\bar{z}} = \langle e^{-i\theta}\rangle + i\delta A = \sum_n \rho_{n,n-1} + i\delta A. \tag{25}$$

For a two−level system with two momentum states $|0\rangle$ and $|-1\rangle$, Eq. (25) is reduced to

$$\frac{dA(\bar{z})}{d\bar{z}} = \rho_{0,-1} + i\delta A. \tag{26}$$

Redefining the off−diagonal element of density matrix as $\acute{\rho}_{0,-1} = \rho_{0,-1}e^{-i\delta\bar{z}} \equiv \acute{B}$ and using the relations $\rho_{0,0} \equiv P_0$ and $\rho_{-1,-1} \equiv P_{-1}$, Eqs. (21)-(23) and Eq. (26) are respectively

$$\frac{dP_0}{d\acute{z}} = -\left(\acute{A}\acute{B}^* + c.c.\right) - DP_0, \tag{27}$$

$$\frac{dP_{-1}}{d\acute{z}} = \left(\acute{A}\acute{B}^* + c.c.\right) + DP_0, \tag{28}$$

$$\frac{d\acute{B}}{d\acute{z}} = -\left(i\acute{\delta} + \frac{D}{2}\right)\acute{B} + \acute{A}(P_0 - P_{-1}), \tag{29}$$

$$\frac{d\acute{A}}{d\acute{z}} = \acute{B}. \tag{30}$$

In deriving Eqs. (27) − (30), we again use the variables $\acute{z} = \sqrt{\bar{\rho}}\bar{z}$ , $\acute{A} = \sqrt{\bar{\rho}}Ae^{-i\delta\bar{z}}$ , $\acute{\delta} = [\delta - (1/2\bar{\rho})]/\sqrt{\bar{\rho}}$, and $D = \beta/\bar{\rho}^{3/2}$. It is obvious that Eqs. (27)−(30) obtained on the basis of the density matrix approach are similar to Eqs. (8)−(11) based on the discrete Wigner model.

The difference between the both models is mainly due to the fact that the Lindblad form of the master equation fulfills the trace−preserving property of the density matrix. On the other hand, the trace of the corresponding components of Wigner function to the diagonal elements of density matrix, $w_0^0$ and $w_{-1}^0$, is not conserved. In the density matrix model, from the second term of Eq. (28) representing the inclusion of the spontaneous emission, the rate of the final state population $dP_{-1}/d\acute{z}$ are only determined by the variation in the initial state population $P_0$, and vice versa as indicated by Eq. (27). In this case, the population can only be exchanged between two momentum eigenstates and no other transitions are allowed. Eqs. (27) and (28) satisfy the trace-preserving condition (i.e., $P_0(\bar{z}) + P_1(\bar{z}) = 1$) as will be demonstrated by numerical examples in the next section. Note that the first terms of Eqs. (27) and (28) correspond to the population change due to the stimulated emission. The opposite sign of these terms ensures the conservation of probabilities due to the stimulated emission.

In the Wigner model, due to the spontaneous emission, $dP_{-1}/d\acute{z}$ is proportional to $P_0 - P_{-1}$ as expressed by the second term on the right-hand side of Eq. (9). In this case, the probability of occupying the final state increases by spontaneous emission from electrons occupying the initial state but decreases by spontaneous emission from electrons occupying the final state. Then, the Wigner model is more exact than the density matrix because the spontaneous emission from the final state is also taken into account. The drawback associated to the trace−preserving property assumed in the density matrix treatment becomes pronounced as the rate of the spontaneous emission increases. As will be shown in the next section, in the practical regime of QFEL operation ($D < 0.07$), the density matrix model is quite applicable where a very good agreement with the discrete Wigner model is observed. It is also noticed that, by comparing Eq. (10) and Eq. (29), the lifetime of coherence in the Wigner model is $1/D$, twice as large as that appeared in the density matrix model. In fact, this does not cause a significant difference between both models since the spontaneous emission rate $D \ll 1/\left(2\bar{\rho}^{3/2}\right)$ in the quantum regime where $\bar{\rho} < 0.4$.

**4. Numerical results and discussion**

In this section, using the density matrix approach, we firstly show the fundamental properties of coherent QFEL emission when the spontaneous emission is negligible (i.e., $D \ll 1$). Next, considering the spontaneous emission, we compare the dynamics of electrons and radiation field predicted by the density matrix and Wigner models using Eqs. (28)−(31) and Eqs. (8)−(11), respectively. The condition for which both models are equivalent is given.

**4.1. Negligible spontaneous emission regime**

Here, we not only address the validity of the density matrix treatment when $D = 0$, but present insights on the electron dynamics described by the density matrix elements.

Assuming $D = 0$, from Eqs. (27) and (28), we get the population difference

$$P_0(\acute{z}) - P_{-1}(\acute{z}) = 1 - 2\left|\acute{A}\right|^2. \quad (31)$$

Then, using Eq. (31) with Eqs. (29) and (30), the evolution of radiation field is described by solving

$$\frac{d^2\acute{A}(\acute{z})}{d\acute{z}^2} + i\acute{\delta}\frac{d\acute{A}(\acute{z})}{d\acute{z}} - \left[1 - 2\left|\acute{A}\right|^2\right]\acute{A} = 0. \quad (32)$$

In Eq. (32), the term $2\left|\acute{A}\right|^2$ is responsible for the nonlinear characteristics of QFEL radiation. The linear regime is dominant if $2\left|\acute{A}\right|^2 \ll 1$ where $P_0(\acute{z}) - P_{-1}(\acute{z}) = 1$. In the linear regime, assuming $\left|\acute{A}\right|$ has a solution in the form of $\left|\acute{A}\right| \propto e^{i\,\acute{\lambda}\acute{z}}$ in Eq. (32) and using the relation $\acute{\delta} = [\delta - (1/2\bar{\rho})]/\sqrt{\bar{\rho}}$, we get

$$\acute{\lambda}^2 + \frac{1}{\sqrt{\bar{\rho}}}[\delta - (1/2\bar{\rho})]\,\acute{\lambda} + 1 = 0. \quad (33)$$

In Fig. (1), using Eq. (33), we plot $\left|\mathrm{Im}(\acute{\lambda})\right|$ vs. $\delta$ at different values of $\bar{\rho}$ in the quantum regime (i.e., $\bar{\rho} = 0.1, 0.2, 0.3,$ and $0.4$). Fig. (1) illustrates the fundamental characteristics of the QFEL operating in the linear regime [3] which are (i) the resonance of the gain occurs at $\delta = 1/2\bar{\rho}$, (ii) the full width of the gain curve is $4\bar{\rho}$, and (iii) the peak of the gain is $\sqrt{\bar{\rho}}$.

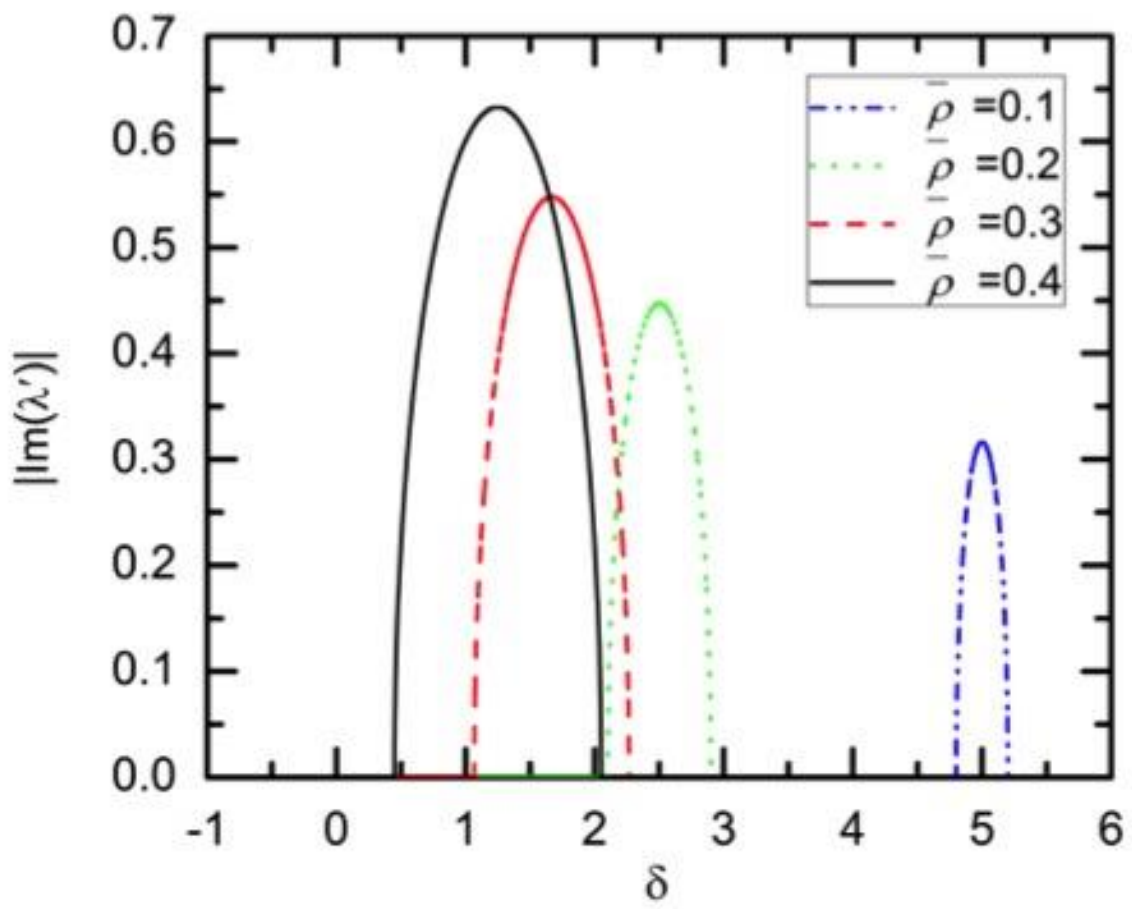

Fig. 1. In the linear regime and when $D = 0$, imaginary part of the complex root of the quadratic equation Eq. (33) vs. $\delta$ for different values of $\bar{\rho}$ in the quantum regime ($\bar{\rho}$= 0.1, 0.2, 0.3, and 0.4).

Using Eq. (33), we show in Fig. 2(a) the evolution of the number of photons per electron $\left|\acute{A}\right|^2$ with $\acute{z}$ in the linear and nonlinear regimes. In these examples, it is assumed that $\bar{\rho} = 0.3$, $\delta = 1/2\bar{\rho}$, and $\acute{B}(0) = 0.01$. In Fig. 2(a), one can realize that the deviation in the results of the linear and nonlinear regimes is small until the first peak. In Fig. 2(a), the position of the first peak is $\acute{z}|_{\mathrm{peak}} \approx 6.0$ which agrees with the predicted value from the relation $\acute{z}|_{\mathrm{peak}} = -\ln\left[\acute{B}(0)/4\right]$ reported in Ref. [8]. Also, in agreement with the results of Ref. [8], Fig. 2(a) shows that the

maximum bunching occurs at $\acute{z} = \acute{z}|_{\text{peak}} \pm 0.88$. In Fig. 2(a), it is seen that the maximum number of photons is 1 in the nonlinear regime.

In Fig. 2(b), we plot $|\acute{A}|^2$, $|\acute{B}|$, and $P_0 - P_{-1}$ vs. $\acute{z}$. From Fig. (2b), it is shown that as $|\acute{A}|^2$ varies periodically from 0 to 1, $P_0 - P_{-1}$ varies from 1 to -1. This behavior can be predicted from Eq. (31). The maximum and minimum values of the population difference, 1 and -1, correspond to the maximum probability of finding an electron in the initial state $n = 0$ and the final state $m = -1$, respectively. Consequently, it is expected that the maximum of the bunching factor $|\acute{B}|$ is 0.5 as shown in Fig. 2(b).

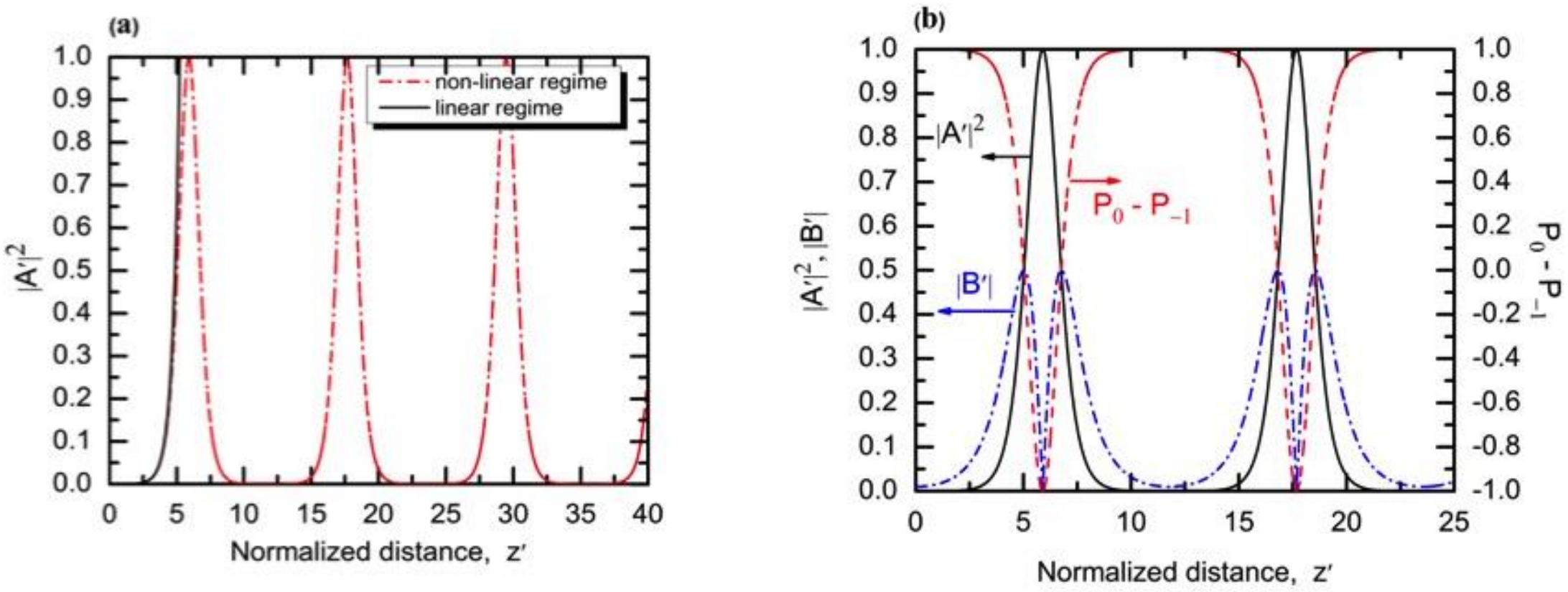

Fig. 2. (a) $|\acute{A}|^2$ vs. $\acute{z}$ in the linear and nonlinear regimes. (b) $|\acute{A}|^2$, $|\acute{B}|$, and $P_0 - P_{-1}$ vs. $\acute{z}$.

**4.2. Non-negligible spontaneous emission regime**

In this section, the QFEL operation is investigated taking into account the inclusion of the spontaneous emission. We present comparisons between the results of the density matrix and those of discrete Wigner model. For this purpose, we solve numerically the set of equations for both models, Eqs. (27)–(30) and Eqs. (8)–(11), respectively.

A comparison between the average number of photons emitted per electron, $|\acute{A}|^2$, vs. $\acute{z}$ for different values of the spontaneous emission rate $D$ is shown in Figs. 3. In these numerical examples, we assume $\bar{\rho} = 0.2$, $\acute{A}(0) = 0$, $\acute{B}(0) = 0.01$, $P_0(0) = 1$, and $P_{-1}(0) = 0$. In Fig. 3(a), identical results for the density matrix and discrete Wigner models are seen when the spontaneous emission is negligible, $D = 0$. In Fig. 3(b), for a moderate spontaneous emission rate $D = 0.05$, the difference between the results of the both models till the first peak of $|\acute{A}|^2$ is very small. From Fig. 3(c), for a large *unfavorable* spontaneous emission rate $D = 0.1$, the results of density matrix deviate significantly from those of the Wigner model. In this regime, the density matrix fails to predict an accurate behavior for the QFEL operation. As discussed above, this is because the trace of the density matrix is conserved by the Lindblad form master equation (Tr($\rho$)=1). Then, the population can only be exchanged between two momentum eigenstates. On the other hand, in the Wigner model, the fact that the electron can emit spontaneously after its first spontaneous transition is taken into account. Therefore, the population inversion in the density matrix model evolves faster than that in the Wigner model. Consequently, the density matrix model predicts stronger effects of spontaneous emission leading to smaller intensity of coherent radiation than that predicted by the Wigner model. To illustrate the latter behavior, in Fig. (4), we plot the trace of probabilities $P_0 + P_{-1}$, the population difference $P_0 - P_{-1}$, and the bunching $\acute{B}$ vs. $\acute{z}$ for $D = 0.05$. In Fig. (4), we assume all parameters as those used in Fig. (3). Fig. 4(a) shows the trace−preserving property is satisfied in the density matrix model, while it is not satisfied in the Wigner model. As seen in Fig. 4(b), in the density matrix model, due to the spontaneous emission of electron, a larger population difference $|P_0 - P_{-1}|$ at shorter interaction distance is observed. Then, the contribution of the coherent emission of electron in the density matrix model is suppressed in a greater way, as shown in Fig. 3(b). Fig. 4(c) shows the bunching factor in the density matrix model is smaller than that in the Wigner model.

In Fig. 5, we plot the first maximum of $|\acute{A}|^2$ as a function of $D$. It can be seen that, in the practical regime when $D \lesssim 0.07$, the results of the density matrix approach are in well agreement with those of the discrete Wigner model. When $D > 0.07$ where the spontaneous emission strongly quenches the coherent lasing process, the density matrix is no longer valid to describe the QFEL interaction. We finally stress on that the regime at which $D > 0.07$ is not a useful operating regime where the coherent radiation is greatly diminished by the spontaneous emission. Therefore, we safely can confirm the density matrix approach is applicable in the practical regime of QFEL operation.

**5. Conclusion**

The density matrix of the Lindblad-type master equation is a powerful tool to describe the quantum FEL interaction. In the quantum regime of the FEL, the diagonal elements of the density matrix fulfill the trace-preserving property where the electron is considered as a two-level system. Then, the exchange of electron populations due to the spontaneous emission occurs only between the initial and final momentum states. The density matrix model is compared with the exact model of the discrete Wigner function in which the electron transition from the final state to a lower momentum state is also considered. We have shown that the results of the density matrix model are in excellent agreement with those of the discrete Wigner model when the rate of the spontaneous emission ensures efficient operation of quantum FELs (when $D \lesssim 0.07$). Then, the approximate model of the density matrix is proved to be rigorous in the practical operating regime of the quantum FEL. It has been shown that the density matrix formalism provides straightforward physical insights into the dynamics of the quantum FEL. However, as the rate of the spontaneous emission increases to a level at which the FEL coherent emission is significantly reduced ($D > 0.07$), the density matrix model becomes invalid.

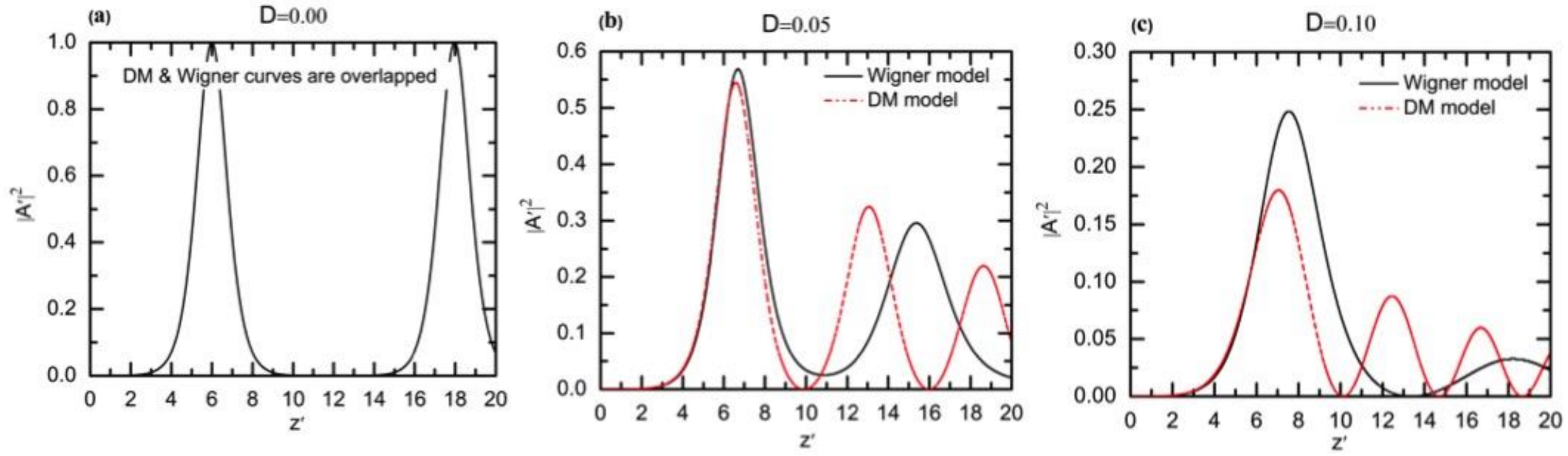

Fig. 3. For the density matrix (DM) and Wigner models, scaled intensity $|\acute{A}|^2$ is plotted as a function of $\acute{z}$ in the quantum regime for different values of the spontaneous emission rate (a) $D = 0.0$ (b) $D = 0.05$ (c) $D = 0.1$.

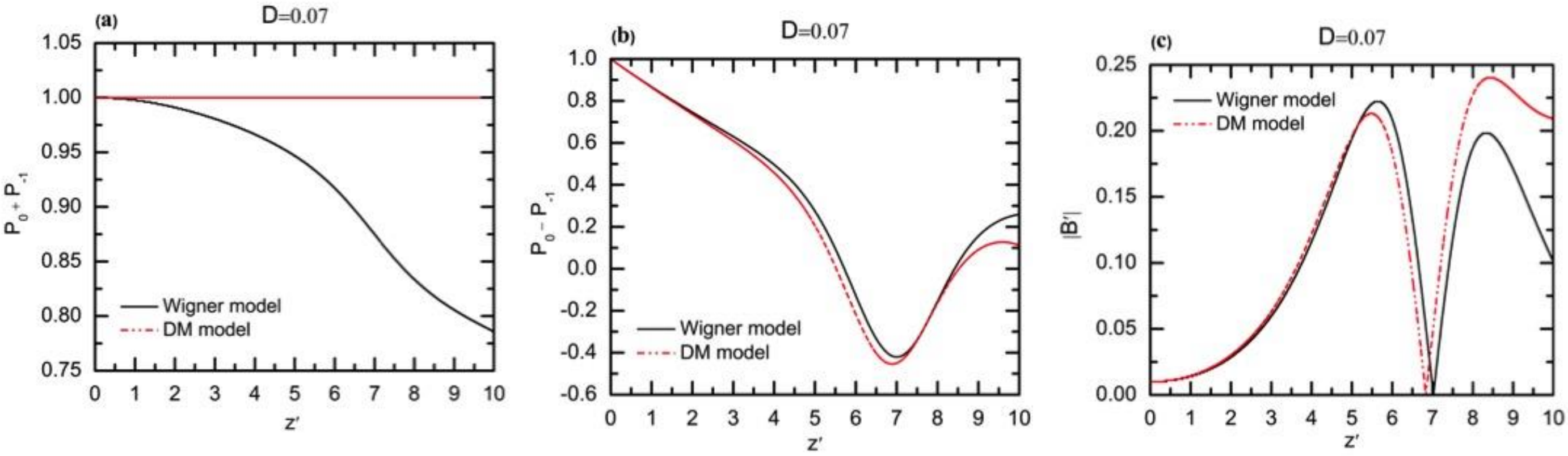

Fig. 4. For the density matrix (DM) and Wigner models when $D = 0.05$, (a) The trace of probabilities $P_0 + P_{-1}$ vs. $\acute{z}$. (b) The population difference $P_0 - P_{-1}$ vs. $\acute{z}$. (c) The bunching factor $\acute{B}$ vs. $\acute{z}$.

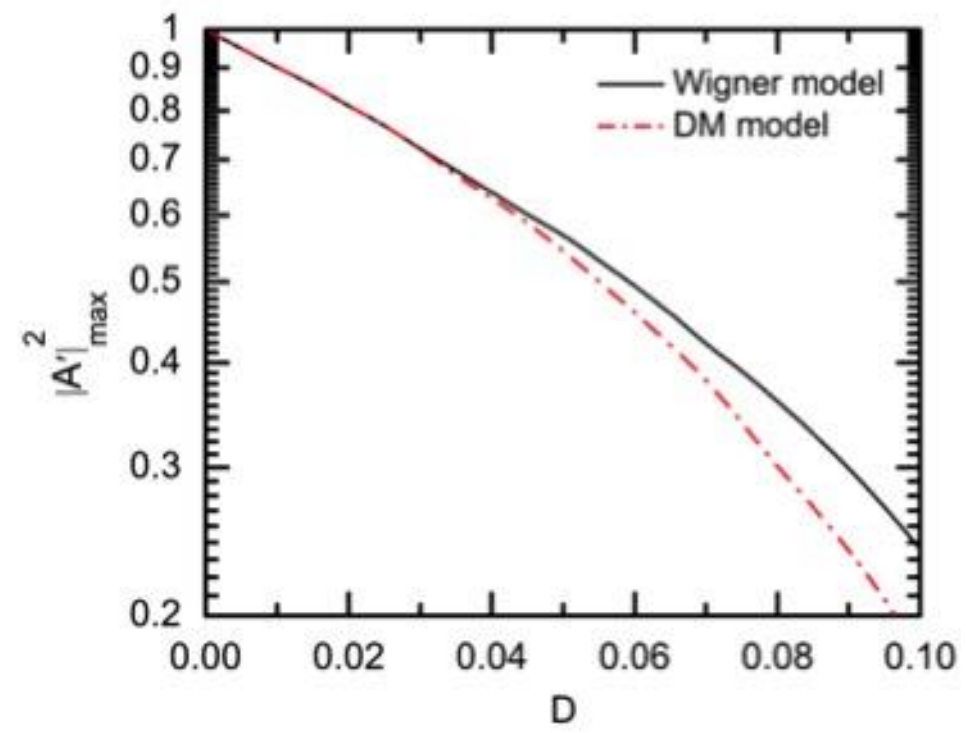

Fig. 5. Comparison between the maximum peak $|\acute{A}|^2$ as predicted by the density matrix (DM) and Wigner models where $|\acute{A}|^2$ is plotted against $D$.

**Acknowledgment**

This work is supported by ASRT-INFN joint project between the Academy of Scientific Research and Technology (ASRT) in Egypt and INFN in Italy.